\documentclass[aps,prb,twocolumn,superscriptaddress,unsortedaddress]{revtex4}

\usepackage{amsmath}
\usepackage{amssymb}
\usepackage{graphicx}
\usepackage{multirow}
\usepackage{array}
\usepackage[english]{babel}

\begin{document}

    \author{G. W. Scheerer}
    \affiliation{Laboratoire National des Champs Magn\'etiques Intenses, CNRS-INSA-UJF-UPS, 143 Avenue de Rangueil, F-31400 Toulouse, France}

    \author{W. Knafo}
    \affiliation{Laboratoire National des Champs Magn\'etiques Intenses, CNRS-INSA-UJF-UPS, 143 Avenue de Rangueil, F-31400 Toulouse, France}

    \author{D. Aoki}
    \affiliation{Institut Nanosciences et Cryog\'enie, SPSMS, CEA-Grenoble, 17 rue des Martyrs, F-38054 Grenoble, France}
    \affiliation{Institute for Materials Research, Tohoku University, Oarai, Ibaraki, 311-1313, Japan}
		
		\author{M. Nardone}
    \affiliation{Laboratoire National des Champs Magn\'etiques Intenses, CNRS-INSA-UJF-UPS, 143 Avenue de Rangueil, F-31400 Toulouse, France}
		
		\author{A. Zitouni}
    \affiliation{Laboratoire National des Champs Magn\'etiques Intenses, CNRS-INSA-UJF-UPS, 143 Avenue de Rangueil, F-31400 Toulouse, France}
		
		\author{J. B\'eard}
    \affiliation{Laboratoire National des Champs Magn\'etiques Intenses, CNRS-INSA-UJF-UPS, 143 Avenue de Rangueil, F-31400 Toulouse, France}
		
		\author{J. Billette}
    \affiliation{Laboratoire National des Champs Magn\'etiques Intenses, CNRS-INSA-UJF-UPS, 143 Avenue de Rangueil, F-31400 Toulouse, France}
		
		\author{J. Barata}
    \affiliation{Laboratoire National des Champs Magn\'etiques Intenses, CNRS-INSA-UJF-UPS, 143 Avenue de Rangueil, F-31400 Toulouse, France}
		
		\author{C. Jaudet}
    \affiliation{Laboratoire National des Champs Magn\'etiques Intenses, CNRS-INSA-UJF-UPS, 143 Avenue de Rangueil, F-31400 Toulouse, France}
		
		\author{M. Suleiman}
    \affiliation{Laboratoire National des Champs Magn\'etiques Intenses, CNRS-INSA-UJF-UPS, 143 Avenue de Rangueil, F-31400 Toulouse, France}
		
		\author{P. Frings}
    \affiliation{Laboratoire National des Champs Magn\'etiques Intenses, CNRS-INSA-UJF-UPS, 143 Avenue de Rangueil, F-31400 Toulouse, France}
		
		\author{L. Drigo}
    \affiliation{Laboratoire National des Champs Magn\'etiques Intenses, CNRS-INSA-UJF-UPS, 143 Avenue de Rangueil, F-31400 Toulouse, France}
		
		\author{A. Audouard}
    \affiliation{Laboratoire National des Champs Magn\'etiques Intenses, CNRS-INSA-UJF-UPS, 143 Avenue de Rangueil, F-31400 Toulouse, France}

		\author{T.D. Matsuda}
    \affiliation{ Advanced Science Research Center, Japan Atomic Energy Agency, Tokai, Ibaraki 319-1195, Japan}

		\author{A. Pourret}
    \affiliation{Institut Nanosciences et Cryog\'enie, SPSMS, CEA-Grenoble, 17 rue des Martyrs, F-38054 Grenoble, France}
		
		\author{G. Knebel}
    \affiliation{Institut Nanosciences et Cryog\'enie, SPSMS, CEA-Grenoble, 17 rue des Martyrs, F-38054 Grenoble, France}

    \author{J. Flouquet}
    \affiliation{Institut Nanosciences et Cryog\'enie, SPSMS, CEA-Grenoble, 17 rue des Martyrs, F-38054 Grenoble, France}

    \title{Fermi surface in the hidden-order state of URu$_2$Si$_2$ under intense pulsed magnetic fields up to 81~T}

    \begin{abstract}
We present measurements of the resistivity $\rho_{x,x}$ of URu$_2$Si$_2$ high-quality single crystals in pulsed high magnetic fields up to 81~T at a temperature of 1.4~K and up to 60~T at temperatures down to 100~mK. For a field \textbf{H} applied along the magnetic easy-axis \textbf{c}, a strong sample-dependence of the low-temperature resistivity in the hidden-order phase is attributed to a high carrier mobility. The interplay between the magnetic and orbital properties is emphasized by the angle-dependence of the phase diagram, where magnetic transition fields and crossover fields related to the Fermi surface properties follow a 1/$\cos\theta$-law, $\theta$ being the angle between \textbf{H} and \textbf{c}. For $\mathbf{H}\parallel\mathbf{c}$, a crossover defined at a kink of $\rho_{x,x}$, as initially reported in [Shishido \textit{et al.}, Phys. Rev. Lett. \textbf{102}, 156403 (2009)], is found to be strongly sample-dependent: its characteristic field $\mu_0H^*$ varies from $\simeq20$~T in our best sample with a residual resistivity ratio $\rm{RRR}=\rho_{x,x}(\rm{300K})/\rho_{x,x}(\rm{2K})$ of $225$ to $\simeq25$~T in a sample with a RRR of $90$. A second crossover is defined at the maximum of $\rho_{x,x}$ at the sample-independent characteristic field $\mu_0H_{\rho,max}^{LT}\simeq30$~T. Fourier analyzes of Shubnikov-de Haas oscillations show that $H_{\rho,max}^{LT}$ coincides with a sudden modification of the Fermi surface, while $H^*$ lies in a regime where the Fermi surface is smoothly modified. For $\mathbf{H}\parallel\mathbf{a}$, i) no phase transition is observed at low temperature and the system remains in the hidden-order phase up to 81~T, ii) quantum oscillations surviving up to 7~K are related to a new and almost-spherical orbit - for the first time observed here - at the frequency $F_\lambda\simeq1400$~T and associated with a low effective mass $m^*_\lambda=(1\pm0.5)\cdot m_0$, where $m_0$ is the free electron mass, and iii) no Fermi surface modification occurs up to 81~T.

    \end{abstract}

\pacs{71.18.+y, 71.27.+a, 75.30.Kz, 72.15.-v}


%
%

    \maketitle

    \section{Introduction}

After more than 20 years of investigations, the heavy-fermion URu$_2$Si$_2$ remains an unsolved issue due to its "hidden-order" phase developing below $T_0=17.5$~K, for which the order parameter has still not been identified \cite{mydosh,palstra85,schablitz,maple,bourdarot,ramirez}. This system is characterized by an Ising anisotropy, with the easy magnetic axis $\textbf{c}$ in the tetragonal structure, resulting in anisotropic electronic properties (magnetic susceptibility \cite{palstra85,ramirez,dawson}, resistivity \cite{palstra86,ohkuni97}, etc.). Superconductivity, whose upper critical field is anisotropic too, sets in below $T_{SC}\simeq1.5$~K.\cite{schablitz,maple,kwok,ohkuni97} Hydrostatic pressure drives the system through a first-order phase transition at $p_c=0.5$~GPa to an antiferromagnetic ground state.\cite{motoyama,amitsuka,hassingerPT} A high magnetic field applied along the $\textbf{c}$-axis also induces a cascade of first-order transitions at the fields $\mu_0 H_1\simeq35$~T, $\mu_0 H_2\simeq37$~T, and $\mu_0 H_3\simeq39$~T, which were probed over the last years using a wide range of experimental techniques: magnetization,\cite{devisser86,sugiyama99, scheerer12} ultrasonic velocity,\cite{suslov,YANAGISAWA} resistivity,\cite{devisser86,kim03,scheerer12} heat capacity,\cite{jaime02} dilatometry,\cite{correa} and thermoelectricity\cite{malone,pourret}. The hidden-order phase is destabilized at $H_1$ and a polarized paramagnetic state is obtained above $H_3$. Between $H_1$ and $H_3$, intermediate magnetic phases are delimited by the critical field $H_2$. As determined recently for Rh-doped URu$_2$Si$_2$, \cite{kuwahara} antiferromagnetic long-range ordering develops in the intermediate phases of URu$_2$Si$_2$ between 35 and 39~T. At low temperature, a maximum of the magnetoresistivity at $\mu_0 H_{\rho,max}^{LT}\simeq30$~T is associated with a Fermi surface modification inside the hidden-order phase.\cite{scheerer12} At high temperature, a crossover leads to maxima at $T_{\rho,max}\simeq40$~K in the electronic, i.e., non-phononic, term of the resistivity and at $T_{\chi,max}\simeq55$~K in the susceptibility, which are related to intersite electronic correlations.\cite{scheerer12} The suppression of these high-temperature scales at 35~T is connected to the destabilization of the hidden-order phase and to the set-up of a high-field polarized regime (see also [\onlinecite{aokiCR}]). When the field rotates from the $\textbf{c}$- to the $\textbf{a}$-axis, the complete phase diagram is pushed towards higher field scales.\cite{sugiyama90,jo,scheererJPSJ}

URu$_2$Si$_2$ is a compensated semi-metal at low temperatures,\cite{kasahara,levallois} for which a sudden reconstruction of the Fermi surface\cite{schoenes,dawson,santander,yoshida10,kawasaki11} occurs at the onset at $T_0$ of the hidden-order phase. Hall effect,\cite{dawson,kasahara} thermoelectric power,\cite{bel} and heat capacity\cite{bel} measurements have further shown that entering in the hidden-order phase induces a strong reduction of the charge carrier number, while an enhanced Nernst effect\cite{bel} and a strong field-induced variation of the resistivity \cite{scheerer12} indicate a highly-increased carrier mobility in the hidden-order phase. For $\mathbf{H} \parallel \mathbf{a}$, a sudden suppression of the field-dependence of the resistivity for $T>T_0$ is due to a significant loss of the carrier mobility.\cite{scheerer12} The Fermi surface of URu$_2$Si$_2$ in its hidden-order phase is partly-known from quantum oscillation experiments, \cite{bergemann,keller,ohkuni97,ohkuni99,hassinger,altara,aoki12} which revealed four Fermi surface sheets associated with the frequencies $F_\eta\simeq93$~T, $F_\gamma\simeq200$~T, $F_\beta\simeq425$~T, and $F_\alpha\simeq1065$~T for $\mathbf{H}\parallel\mathbf{c}$. The Sommerfeld coefficient $\gamma_{FS}\simeq 37.5$~mJ/mol$\cdot$K$^2$ estimated from these Fermi surface measurements \cite{hassinger} corresponds to 55~\% of the Sommerfeld coefficient $\gamma_{Cp}\simeq 65$~mJ/mol$\cdot$K$^2$ extracted from specific heat data.\cite{maple} In the light of band structure calculations, it is not clear whether an electron or a hole Fermi surface is missing from the quantum oscillations experiments: while Oppeneer \textit{et al.} \cite{oppeneer11} find that a large hole Fermi surface is missing, Ikeda \textit{et al.} \cite{ikeda12} find out that a heavy-electron Fermi surface is missing, in accordance with charge and transport measurements. Recently, a heavy-electron Fermi surface branch has been reported from cyclotron resonance experiments,\cite{tonegawa,tonegawa2013} and was estimated to account for almost 30~$\%$ of the Sommerfeld coefficient determined from the specific heat, suggesting that only 20$\%$ of the Fermi surface would remain unknown. In a high field applied along \textbf{c}, Fermi surface modifications within the hidden-order phase, i.e., up to $\mu_0H_1=35$~T, have been reported in Shubnikov-de Haas (SdH) oscillations spectra,\cite{shishido,aoki12,altara,jo} while the Fermi surface in the polarized regime above $\mu_0H_3=39$~T has not yet been determined. We note that slight discrepancies are found between the Fermi surface frequencies extracted from the different sets of high-field quantum oscillations measurements.\cite{shishido,aoki12,altara,jo} In particular, the new frequency $F_\varepsilon\simeq1300$~T associated with a light mass $m_\varepsilon=2.7m_0$ reported in Hall resistivity above 20~T by Shishido \textit{et al.}\cite{shishido} has not been reproduced yet.

We present here a study of the resistivity of high-quality URu$_2$Si$_2$ single crystals in high magnetic fields up to 60~T at temperatures down to 100~mK and in fields up to 81~T at 1.4~K. Experimental details are given in Section II. In Section III, measurements with the magnetic field applied along the magnetic easy axis \textbf{c} are presented. The strong sample-dependence of the magnetoresistivity in the hidden-order state is characterized carefully. A widespread study of the resistivity in various configurations is presented in Section IV: the effect of a magnetic field rotating in the (\textbf{a},\textbf{c}) and (\textbf{a},\textbf{a}) planes is investigated for both transverse and longitudinal configurations (electrical contacts perpendicular or parallel, respectively, to the field direction). The angle-dependence of the phase transitions $H_1$, $H_2$, $H_3$, and crossovers $H^*$ and $H_{\rho,max}^{LT}$ is presented. In Section V, Shubnikov-de Haas oscillations are investigated and permit to probe the high-field Fermi surface. For $\mathbf{H}\parallel\mathbf{c}$, magnetic-field-induced Fermi surface modifications are observed inside the hidden-order phase. For $\mathbf{H}\parallel\mathbf{a}$ quantum oscillations from the branches $\gamma$ and $\alpha$, and from a new light-mass branch $\lambda$ are observed; they indicate that the Fermi surface is not modified in magnetic fields up to 81~T. By extending our work published in Ref. [\onlinecite{scheerer12}] to lower temperatures, higher fields, and new field orientations, this study provides further evidences of the interplay between magnetism, Fermi surface reconstructions, and the hidden-order in URu$_2$Si$_2$.


\section{EXPERIMENTAL DETAILS}

We have measured the resistivity of two high-quality single crystals of URu$_2$Si$_2$ grown by the Czochralski technique in a tetra-arc furnace. Details about the crystal growth can be found in Ref. [\onlinecite{aoki2010}]. The high-field electrical resistivity $\rho_{x,x}$ was investigated by the four-contact method using the lock-in technique, at frequencies from 40 to 70~kHz. The electric current $I$ and voltage $U$ have been applied and measured, respectively, along the [100] direction. The residual resistivity ratio $\rm{RRR}=\rho_{x,x}(\rm{300K})/\rho_{x,x}(\rm{2K})$ defined at zero-field reaches $\simeq90$ for sample $\sharp1$ and $\simeq225$ for sample $\sharp2$, indicating their high quality (see Ref. [\onlinecite{matsuda}] for a study of the sample-dependence of URu$_2$Si$_2$ single crystals properties). Pulsed magnetic field experiments were done at the Laboratoire National des Champs Magn\'etiques Intenses of Toulouse (LNCMI-T), France. Pulsed fields have been generated either by 6-mm-bore 60-T or 70-T magnets with a pulse duration of 150~ms, or by 20-mm-bore 60-T magnets with a pulse duration of 300~ms. A magnetic field up to 81~T has been generated by a new double coil,\cite{beard} made of an outer coil delivering a long pulse of 250~ms up to 30~T and an inner coil delivering a short pulse of 75~ms from 30~T to 81~T, which allows a unique duration of the pulse of 10.2~ms above 70~T.\cite{beard} Standard $^4$He cryostats, as well as a home-made non-metallic $^3$He-$^4$He-dilution fridge specially designed for the pulsed magnetic fields have been used to reach temperatures down to 1.4~K and 100~mK, in magnetic fields up to 81~T and 60~T, respectively. Electrical transport probes with static or rotating sample support have been used to study the samples properties in different configurations of the field: the transverse configurations ($\mathbf{H}\parallel\mathbf{c}$; $\mathbf{I},\mathbf{U}\perp\mathbf{H}$) and ($\mathbf{H}\parallel\mathbf{a}$; $\mathbf{I},\mathbf{U}\perp\mathbf{H}$) have been probed using a static support, while configurations with \textbf{H} applied along various directions in the (\textbf{a},\textbf{c}) and (\textbf{a},\textbf{a}) planes have been investigated using a rotation probe. Complementarily, the resistivity of sample $\sharp2$ has been studied at $T=32$~mK in a transversal configuration for $\mu_0 H$ up to 13~T rotating in the (\textbf{a},\textbf{c}) plane. The Shubnikov-de Haas oscillations and their frequencies extracted from this "low-field" experiment - not shown here - are in perfect agreement with those published in [\onlinecite{hassinger}].


\section{SAMPLE-DEPENDENCE OF THE HIGH-FIELD resistivity}

\begin{figure}[t]
\centering
\includegraphics[width=8cm]{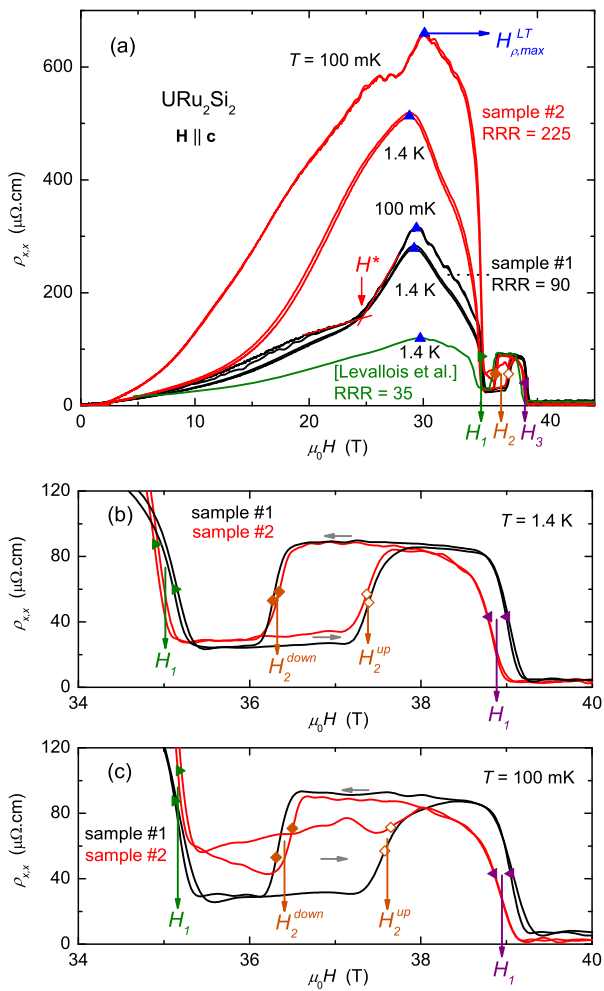}
\caption{(a) Transverse resistivity $\rho_{x,x}$ versus $\textbf{H}$ applied along \textbf{c} of samples $\sharp1$ and $\sharp2$ at $T=100$~mK and $T=1.4$~K, and of a third sample measured by Levallois \textit{et al}. \cite{levallois} at $T=1.4$~K. Zoom on $\rho_{x,x}(H)$ of samples $\sharp1$ and $\sharp2$ for $\mathbf{H}\parallel\mathbf{c}$ between 34~T and 40~T at (b) $T=1.4$~K and (c) $T=100$~mK. The grey arrows indicate the rise and fall of the pulsed field.}
\label{RRR1}
\end{figure}

Figure \ref{RRR1}(a) presents, at $T=100$~mK and 1.4~K, the transverse resistivity $\rho_{x,x}$ of two URu$_2$Si$_2$ samples of different qualities [samples $\sharp1$ (RRR = 90) and $\sharp2$ (RRR = 225) measured here and a third sample (RRR = 35) measured by Levallois \textit{et al}.\cite{levallois} versus a magnetic field applied along the \textbf{c}-axis. This plot extends to sub-kelvin temperature the study performed on samples $\sharp1$ and $\sharp2$ above 1.4 K in Ref. [\onlinecite{scheerer12}], where the ($H$,$T$) phase diagram has been extracted from resistivity and magnetization data. Superconductivity develops below $T_{sc}=1.5$~K and leads to $\rho_{x,x}=0$ for $\mu_0H<\mu_0 H_{c2}\simeq2.5$~T at $T=100$~mK. Figures \ref{RRR1}(b) and \ref{RRR1}(c) focus on the resistivity of samples $\sharp1$ and $\sharp2$ at $T=1.4$~K and 100~mK, respectively, in the field range $34\leq\mu_0H\leq40$~T. At $T=1.4$~K, $\rho_{x,x}$ is almost sample-independent for $\mu_0H>\mu_0H_1=35$~T and sharp steps are observed at the first-order transition fields $H_1$, $H_2$ and $H_3$. However, at $T=100$~mK $\rho_{x,x}$ becomes sample-dependent also in the field range $H>H_1$. While $\rho_{x,x}$ of sample $\sharp1$ is almost the same at 100~mK as at 1.4~K, $\rho_{x,x}$ of sample $\sharp2$ is strongly modified at 100~mK, the transition fields $H_1$, $H_2$ and $H_3$ being more difficult to define, in particular for increasing field, than at 1.4~K. Knowing that sample $\sharp2$ has the highest RRR, and thus the highest electronic mean free path, this result indicates an interplay between the cyclotron motion of the electrons and their scattering on the magnetic ions.

\begin{widetext}

\begin{figure}[h]
\centering
\includegraphics[width=16.5cm]{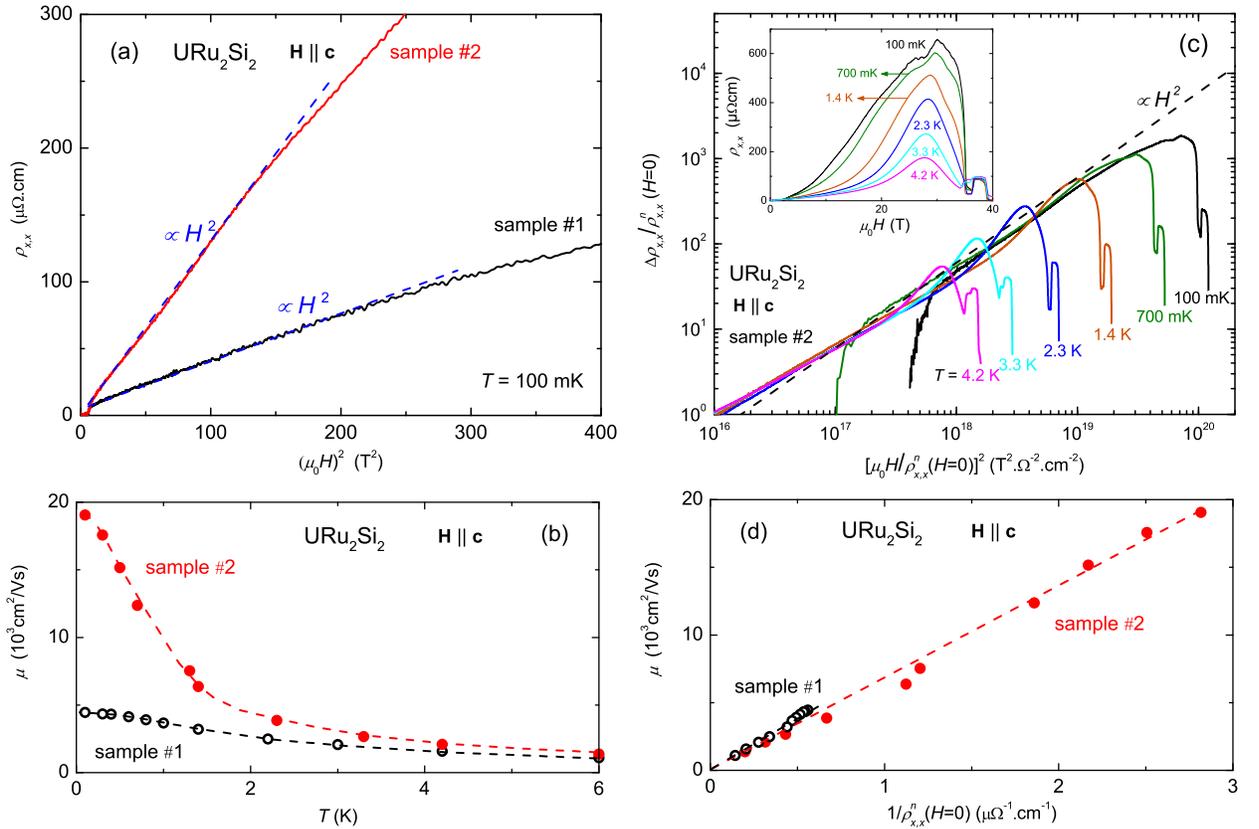}
\caption{(a) Plot of the transverse resistivity $\rho_{x,x}$ versus $H^2$, for $\mu_0H<20$~T, for  samples $\sharp1$ and $\sharp2$ at $T=100$~mK.
(b) Plot of the mobility $\mu$ versus $T$ for samples $\sharp1$ and $\sharp2$ at 100~mK~$\leq T \leq1.4$~K. (c) Kohler plot of $\Delta\rho_{x,x}/\rho^n_{x,x}(H=0)$ versus $[\mu_0H/\rho^n_{x,x}(H=0)]^2$ in a log-log scale for sample $\sharp2$ at temperatures from 100~mK to 4.2~K. (d) Plot of the mobility $\mu$ versus $1/\rho^n_{x,x}(H=0)$ for samples $\sharp1$ and $\sharp2$ at 100~mK~$\leq T \leq1.4$~K. The dashed lines are guides to the eyes.}
\label{RRR2}
\end{figure}

\end{widetext}

In a compensated electron-hole two-band picture, the field-induced variation of the resistivity can be approximated at low fields by:
\begin{eqnarray}
\Delta\rho(H)/\rho(H=0)=\mu_e\mu_h(\mu_0 H)^2,
    \label{rho_comp}
\end{eqnarray}
where $\mu_e$ and $\mu_h$ are the electron and hole mobilities, respectively.\cite{pippard,ziman,scheerer_thesis} As shown in Figure \ref{RRR2}(a), the resistivities $\rho_{x,x}$ of samples $\sharp1$ and $\sharp2$ show, in agreement with previous studies\cite{levallois,kasahara}, a field-dependence close to $\propto H^2$ at $T=100$ mK for $\mu_0H_{c,2}\simeq2.5$~T~$<\mu_0H<10$~T. Clear deviations from a $H^2$ behavior are observed at higher fields, being presumably related to the complex multi-band structure of the Fermi surface\cite{oppeneer11,ikeda12}. In the following we consider the zero-field value $\rho^n_{x,x}(H=0)$ and the field-dependent term $\Delta\rho_{x,x}=\rho_{x,x}-\rho^n_{x,x}(H=0)$ of the resistivity in the normal non-superconducting state: below $T_{sc}=1.5$~K, $\rho^n_{x,x}(H=0)$ of a virtual normal state was estimated by extrapolating to the low temperatures a $T^{1.5}$ law preliminarily fitted to $\rho_{x,x}(T)$ above $T_{sc}=1.5$~K (see also Ref. [\onlinecite{matsuda}]). From fits of $\Delta\rho_{x,x}/\rho^n_{x,x}(H=0)$ versus $H^2$ to Equation \ref{rho_comp}, we extract the mobility averaged over the different bands $\mu=\langle\sqrt{\mu_e\mu_h}\rangle$, which reaches $4.5\times10^3$~and $1.9\times10^4$~cm$^{2}$/Vs at $T=100$~mK for samples $\sharp1$ and $\sharp2$, respectively. A plot of $\mu$ versus $T$ up to 6~K is shown in Figure \ref{RRR2}(b) for samples $\sharp1$ and $\sharp2$ and illustrates that the carrier mobility is enhanced as the temperature is decreased and the sample quality is increased (sample $\sharp2$ has a much higher quality, as indicated by its RRR, than sample $\sharp1$). As shown in Figure \ref{RRR1}(a), a maximum of $\rho_{x,x}$ is obtained at $\mu_0H_{\rho,max}^{LT}\simeq30$~T, i.e., inside the hidden-order phase, for all the samples. At $T=1.4$~K, $\rho_{x,x}$ of sample $\sharp2$ reaches $\simeq500$~$\mu\Omega.$cm at $H_{\rho,max}^{LT}$, which is twice the value of $\rho_{x,x}(H_{\rho,max}^{LT})$ of sample $\sharp1$ ($\simeq300$~$\mu\Omega.$cm) and five times that of the third sample studied by Levallois \textit{et al.} ($\simeq100$~$\mu\Omega.$cm).\cite{levallois} At $T=100$~mK, $\rho_{x,x}(H_{\rho,max}^{LT})$ of sample $\sharp1$ increases slightly compared to its value at 1.4~K, while $\rho_{x,x}(H_{\rho,max}^{LT})$ of sample $\sharp2$ increases significantly, reaching $\simeq650$~$\mu\Omega.$cm. The maximum at $H_{\rho,max}^{LT}$ indicates a crossover within the hidden-order phase between a low-field Fermi surface with a high carrier mobility to a high-field Fermi surface with a low carrier-mobility. The higher the RRR, the higher is $\rho_{x,x}(H^{LT}_{\rho,max})$, confirming that the transverse resistivity in the hidden-order phase and its broad maximum at 30 T are dominated by an orbital contribution, i.e., the field-induced cyclotron motion of the charge carriers.\cite{scheerer12}

Figure \ref{RRR2}(c) presents in a log-log scale a Kohler plot, i.e., a plot of $\Delta\rho_{x,x}/\rho^n_{x,x}(H=0)$ versus $[\mu_0H/\rho^n_{x,x}(H=0)]^2$, for sample $\sharp2$ at temperatures from 100~mK to 4.2~K. The raw magnetoresistivity $\rho_{x,x}$ versus field data used for the Kohler plot are shown in the Inset of Figure \ref{RRR2}(c). In the Kohler plot, all data sets fall on a single curve, whose field-dependence is close to $\propto H^2$, at fields smaller than 20~T, indicating that a single relaxation time $\tau$ can describe the different bands responsible for the high magnetoresistivity \cite{ziman,scheerer_thesis}. Above 20~T, deviations due to Fermi surface reconstructions are observed. Figure \ref{RRR2}(d) shows that a plot of the mobility $\mu$, extracted here at different temperatures for samples $\sharp1$ and $\sharp2$, versus $1/\rho^n_{x,x}(H=0)$ coincides with a linear function independent of the sample quality. This indicates that the sample- and temperature-dependences of the relaxation time $\tau$ drive those of both $\mu$ and $\rho^n_{x,x}(H=0)$, with a relationship $\mu=a/\rho^n_{x,x}(H=0)=b\tau$, where $a$ and $b$ are constants independent of the temperature and of the sample quality. Despite the complex multi-band structure of the Fermi surface of URu$_2$Si$_2$, \cite{hassinger,oppeneer11,ikeda12,tonegawa,tonegawa2013} its transverse magnetoresistivity in high fields $\mathbf{H}\parallel\mathbf{c}$ can thus be rather-well described, in a first approximation, by a simple compensated electron-hole two-band picture where the average mobility and the zero-field resistivity are simply controlled by a unique relaxation time $\tau$.

Fig. \ref{RRR1}(a) also shows that the resistivity $\rho_{x,x}(H)$ of sample $\sharp1$ exhibits an inflexion point followed by a sudden increase of slope at $\mu_0 H^*=24.7\pm0.5$~T at $T=100$~mK and $\mu_0 H^*=24.6\pm0.8$~T at $T=1.4$~K, i.e., well below $H_{\rho,max}^{LT}$. For sample $\sharp2$ at $T=1.4$~K, such wave-like anomaly is not observed in $\rho_{x,x}(H)$ at fields smaller than $H_{\rho,max}^{LT}$ and $H^*$ cannot be defined. In Section IV, $\mu_0H^*\simeq20$~T is extrapolated for sample $\sharp2$ in $\mathbf{H}\parallel\mathbf{c}$ from transverse resistivity measurements in a field rotating from $c$ to $a$. Another kink developing at around 27~T in the resistivity of sample $\sharp2$ at $T=100$~mK may be related to a low-frequency quantum oscillation. For $\mathbf{H}\parallel\mathbf{c}$, an anomaly in the resistivity or Hall effect similar to that observed in the resistivity of sample $\sharp1$ at $\mu_0 H^*\approx25$~T was observed at subkelvin temperatures by Shishido \textit{et al}.\cite{shishido}, Altarawneh \textit{et al}.\cite{altara}, and Aoki \textit{et al}.\cite{aoki12} at $\mu_0 H^*\simeq22.5$, 24, and 24~T, respectively, but not by Levallois \textit{et al}.\cite{levallois} at $T=1.4$~K. In Refs. [\onlinecite{shishido,altara,aoki12}] the anomaly at $H^*$ has been further related to a field-induced Fermi surface modification, as revealed by changes of the Shubnikov-de Haas frequencies. In Section V, we discuss the relationship between $H^*$ and $H_{\rho,max}^{LT}$ to field-induced Fermi surface modifications.

\section{ANGULAR DEPENDENCE OF THE HIGH-FIELD resistivity}

URu$_2$Si$_2$ exhibits highly anisotropic bulk properties related to its crystal structure. For instance, the in-plane resistivity $\rho_{x,x}(T)$ is twice bigger than the out-of-plane resistivity $\rho_{z,z}(T)$,\cite{palstra86} the superconducting critical field $\mu_0 H_{c,2}$ reaches $\simeq2-3$~T for $\mathbf{H}\parallel \mathbf{c}$ and $\simeq10-13$~T for $\mathbf{H}\parallel \mathbf{a}$,\cite{kwok,ohkuni97,brison95} and the hidden-order phase is destabilized at 35~T for $\mathbf{H}\parallel \mathbf{c}$, while no field-induced transition is observed up to 81~T for $\mathbf{H}\parallel \mathbf{a}$ (see Section V). Sugiyama \textit{et al.}\cite{sugiyama90} and Jo \textit{et al.}\cite{jo} have shown, by measuring the magnetization and the resistivity, respectively, that $H_1$, $H_2$, and $H_3$ all follow a $1/cos\theta$ law, where $\theta$ is the angle between \textbf{c} and \textbf{a}. Shishido \textit{et al.}\cite{shishido} and Aoki \textit{et al.}\cite{aoki12} observed that the crossover field $H^*$ in the resistivity is governed by a $1/cos\theta$ law as well. In Ref. [\onlinecite{scheererJPSJ}], we have shown that the field $H_{\rho,max}^{LT}$ at the maximum of $\rho_{x,x}(H)$ also follows a $1/cos\theta$ law. We extend here the study of the angle-dependence of the high-field resistivity of URu$_2$Si$_2$ by a systematic investigation of $\rho_{x,x}(H)$ for $\mu_0H$ up to 60~T applied in the three main planes i) (\textbf{c},\textbf{a}) with $\mathbf{a}\parallel\mathbf{I},\mathbf{U}$, ii) (\textbf{c},\textbf{a}) with $\mathbf{a}\perp\mathbf{I},\mathbf{U}$, and iii) (\textbf{a},\textbf{a}). Our two samples $\sharp1$ and $\sharp2$ have been characterized in rotating fields.

\begin{figure}
\centering
\includegraphics[width=8cm]{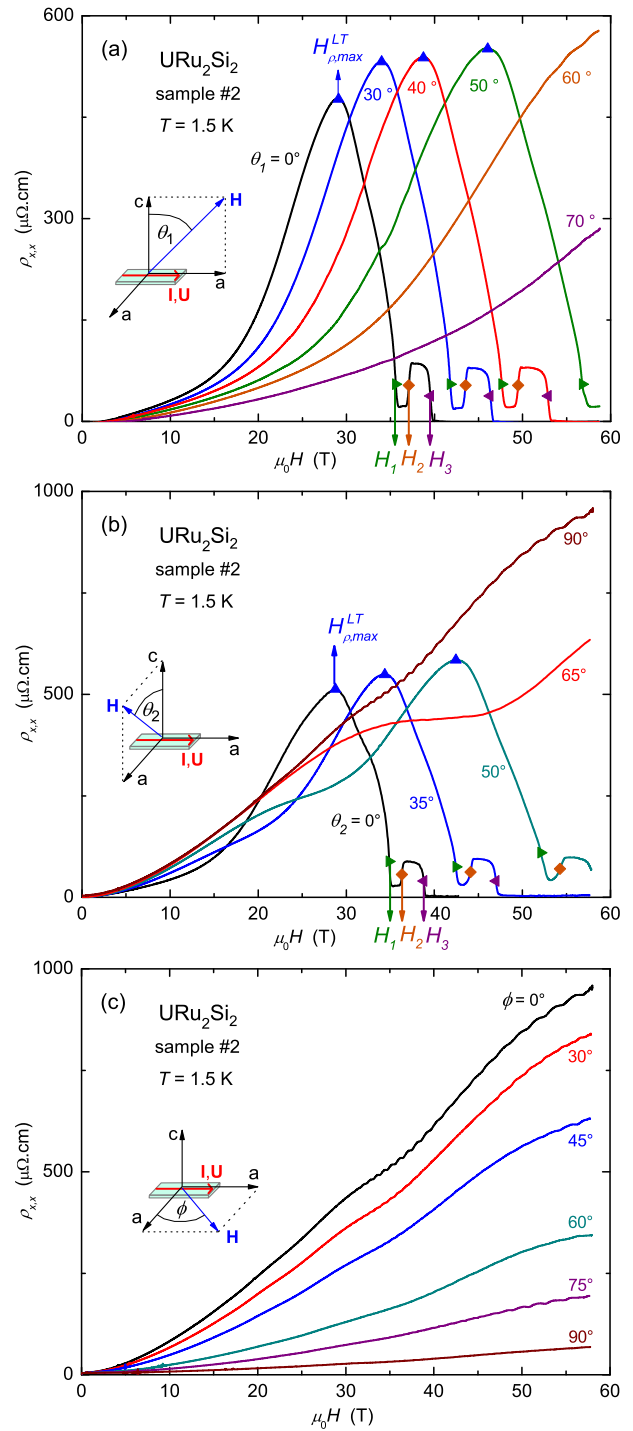}
\caption{
(a) Resistivity $\rho_{x,x}$ versus $H$ of sample $\sharp2$ at $T=1.5$~K for different angles $\theta_1$ between $\mathbf{H}$ and $\mathbf{c}$. The field is turning from the transverse ($\theta_1$ = 0) to the longitudinal ($\theta_1$ = 90$^\circ$) configurations. (b) Resistivity $\rho_{x,x}$ versus $H$ of sample $\sharp2$ at $T=1.6$~K for different angles $\theta_2$ between $\mathbf{H}$ and $\mathbf{c}$. The magnetic field is turning in the transverse plane. (c) Resistivity $\rho_{x,x}$ versus $H$ of sample $\sharp2$ at $T=1.5$~K for different angles $\phi$ between $\mathbf{H}$ and $\mathbf{a}$. The field is turning in the (\textbf{a},\textbf{a}) plane from the transverse ($\phi=0$) to the longitudinal ($\phi=90^\circ$) configurations.
}
\label{figROT1}
\end{figure}

\begin{figure}[t]
\centering
\includegraphics[width=8cm]{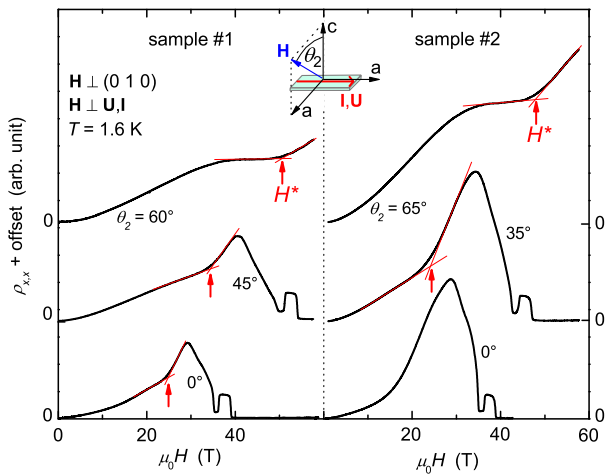}
\caption{
Focus on the anomaly at $H^*$ in the resistivity of samples $\sharp1$ and $\sharp2$ for different orientations of \textbf{H} in the transverse (\textbf{a},\textbf{c}) plane.
}
\label{figROT2}
\end{figure}

\begin{figure}[t]
\centering
\includegraphics[width=8cm]{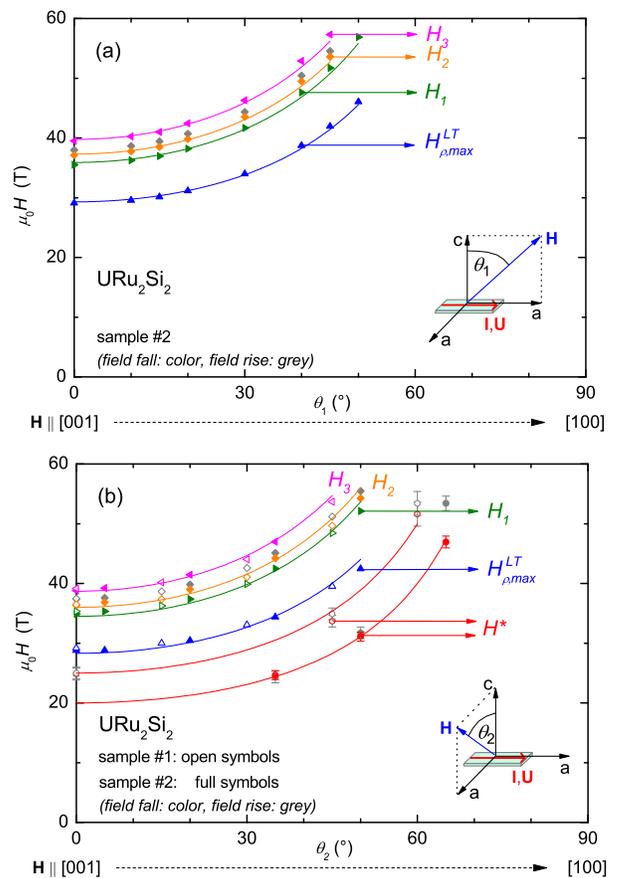}
\caption{
Angle dependence of the transition fields $H_1$, $H_2$, and $H_3$ and crossover fields $H^{LT}_{\rho,max}$ and $H^*$ of sample $\sharp1$ (open symbols) and sample $\sharp2$ (closed symbols). Data from the rise of the pulse are in grey, data from the fall of the pulse are in color. The solid lines represent 1/$\cos\theta$-fits to the data.
}
\label{figROT3}
\end{figure}

Figure \ref{figROT1}(a) shows the resistivity of sample $\sharp2$ at $T\simeq1.5$~K for different angles $\theta_1$ between the magnetic field $\mathbf{H}$ and the \textbf{c}-axis. The magnetic field is turning from the transverse ($\mathbf{H} \parallel \mathbf{c}$; $\mathbf{H} \perp \mathbf{I},\mathbf{U}$; $\theta_1$ = 0$^\circ$) to the longitudinal ($\mathbf{H} \parallel \mathbf{a}$; $\mathbf{H} \parallel \mathbf{I},\mathbf{U}$; $\theta_1$ = 90$^\circ$) configurations, as illustrated by insets to the graphs. When $\theta_1$ increases, the general form of the resistivity remains unchanged, but the anomalies are shifted to higher field values. The maximal value of $\rho_{x,x}$ at $H^{LT}_{\rho,max}$ is also slightly increasing with $\theta_1$. Figure \ref{figROT1}(b) shows the resistivity of sample $\sharp2$ for different angles $\theta_2$ between \textbf{H} and \textbf{c}, where \textbf{H} lies in the (\textbf{a},\textbf{c}) plane perpendicular to the electric current, and rotates from the ($\mathbf{H} \parallel \mathbf{c}$; $\mathbf{H} \perp \mathbf{I},\mathbf{U}$; $\theta_2$ = 0$^\circ$) to the ($\mathbf{H} \parallel \mathbf{a}$; $\mathbf{H} \perp \mathbf{I},\mathbf{U}$; $\theta_2$ = 90$^\circ$) transverse configurations. Again, the fields $H_1$, $H_2$, $H_3$, and $H^{LT}_{\rho,max}$ shift to higher field values with increasing angle $\theta_2$. Remarkably, the heights of the plateaus between $H_1$ and $H_3$ are independent of the orientation of the magnetic field relatively to the \textbf{c}-axis or to the current [cf. Fig. \ref{figROT1}(a,b)] and of the sample quality (see Fig. \ref{RRR1}). In Figure \ref{RRR1}(a) (see Sec. III), a kink was observed at $\mu_0H^* =25$~T in the resistivity of sample $\sharp1$, but not in the resistivity of sample $\sharp2$, for $\mathbf{H} \parallel \mathbf{c}$. In the transverse-to-transverse rotation configuration [see Figs. \ref{figROT1}(b) and \ref{figROT2}], the anomaly at $H^*$ is unveiled in sample $\sharp2$ for $\theta_2\geq35^\circ$, showing a $\theta_2$-dependence similar than that of the anomaly at $H^*$ in sample $\sharp1$. For $\theta_2=90^\circ$, i.e., for $\mathbf{H} \parallel \mathbf{a}$, the transverse resistivity increases continuously up to the highest applied field and no field-induced transition or crossover is observed. Quantum oscillations, whose analysis is given in Section V, are discernable in the high-field magnetoresistivity.

Figure \ref{figROT3} shows the angle-dependence of the transition fields $H_1$, $H_2$, and $H_3$, and the crossover fields $H^{LT}_{\rho,max}$ and $H^*$. Slight misalignments of the samples in the magnetic field are responsible for small differences between the plots in the (a) and (b) panels, which correspond to the transverse-to-longitudinal [Figure \ref{figROT1}(a)] and transverse-to-transverse [Figure \ref{figROT1}(b)] configurations, respectively. The transition fields $H_1$, $H_2$ and $H_3$ are related to the \textit{f}-electron magnetic properties and all follow a 1/$\cos\theta$-law. Their angle-dependence is a direct consequence of the strong Ising-character of the magnetic anisotropy. For both samples $\sharp1$ and $\sharp2$, the crossover fields $H^{LT}_{\rho,max}$ and $H^*$ related to Fermi surface modifications (cf. Section V) show the same 1/$\cos\theta$ dependence as that of the magnetic transition fields. These Fermi surface modifications are thus controlled by the projection of the field along the easy magnetic axis \textbf{c}, which illustrates the interplay between the component of the magnetization along \textbf{c} and the Fermi surface in URu$_2$Si$_2$. A fit by a $1/cos\theta$ law allows extracting $\mu_0 H^*\simeq20$~T for sample $\sharp2$ in the limit of $\theta_2\simeq0^\circ$, i.e., for $\mathbf{H} \parallel \mathbf{c}$. $\mu_0 H^*$ in sample $\sharp2$ is much smaller than the values of 25~T found for sample $\sharp1$ and those between 22.5 and 24~T reported in the literature. \cite{shishido,altara,aoki12} For $\mathbf{H} \parallel \mathbf{c}$, the anomaly at $H^*$ in the resistivity of sample $\sharp2$ might be hidden by an additional orbital contribution whose intensity decreases at high $\theta_2$ angles.

Figure \ref{figROT1}(c) shows the resistivity of URu$_2$Si$_2$ for different directions of the magnetic field inside the (\textbf{a},\textbf{a}) plane, $\phi$ being the angle between the magnetic field \textbf{H} and the \textbf{a}-axis. The transverse ($\phi=0^\circ$) to longitudinal ($\phi=90^\circ$) configurations are explored. The curves show Shubnikov-de Haas quantum oscillations, which are analyzed in Section V. The resistivity decreases with increasing angle $\phi$ and the field-dependent term vanishes almost totally at $90^\circ$. In Ref. [\onlinecite{scheerer12}], we have shown that the strong field-induced transverse resistivity $\rho_{xx}(H)$, which develops below $T_0$ for $\mathbf{H} \parallel \mathbf{a}$, is characteristic of the hidden-order phase. The fact that this contribution vanishes in the longitudinal configuration for $\mathbf{H} \parallel \mathbf{a}\parallel\mathbf{I},\mathbf{U}$, confirms the orbital origin of $\rho_{x,x}$ for $\mathbf{H} \parallel \mathbf{a}\perp\mathbf{I},\mathbf{U}$.

	
\section{HIGH-FIELD FERMI SURFACE}

\subsection{Quantum oscillations for $\mathbf{H}\parallel\mathbf{c}$}

\begin{figure}[b]
\centering
\includegraphics[width=8.7cm]{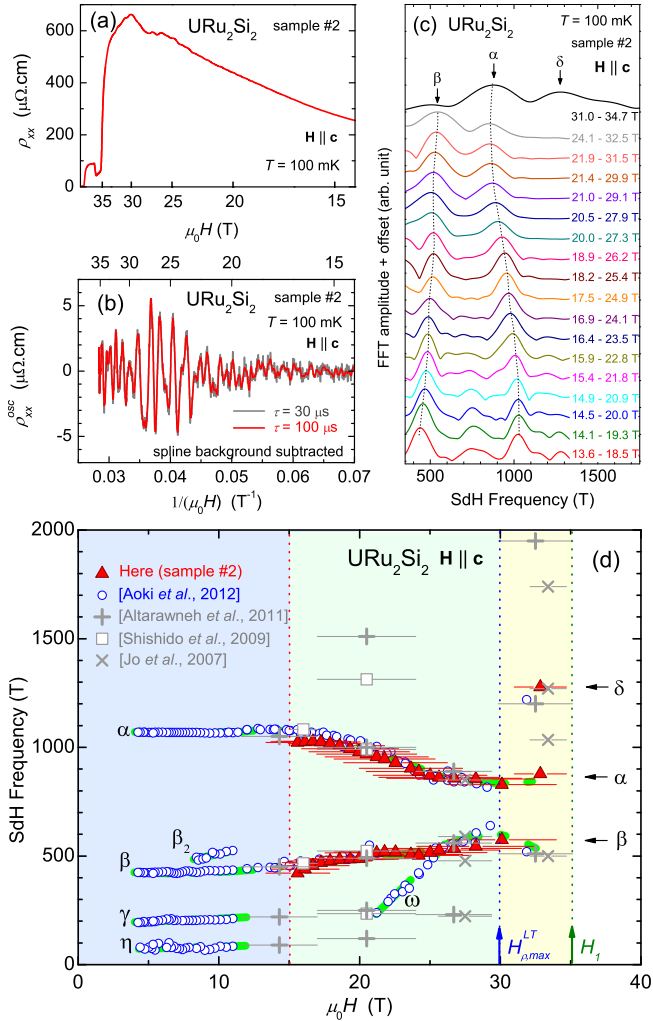}
\caption{(a) Resistivity $\rho_{x,x}$ of sample $\sharp2$ at $T=100$~mK in $\mathbf{H}\parallel\mathbf{c}$. (b) Oscillating signal extracted using a spline background, with the time constants $\tau=30$~and~100~$\mu$s (digital lock-in). (c) Fourier transform spectra for a large set of small field windows. (d) Field-dependence of the Shubnikov-de Haas frequencies extracted here (pulsed fields) and compiled from Refs. [\onlinecite{altara,jo,aoki12,shishido}] (steady fields). The horizontal bars indicate the field windows of the Fourier transform and the dotted lines are guides to the eyes.
}
\label{QOHc2}
\end{figure}

Figure \ref{QOHc2}(a) shows the resistivity $\rho_{x,x}$ of sample $\sharp2$ at $T=100$~mK in a pulsed magnetic field $\mathbf{H}\parallel\mathbf{c}$, and  Figure \ref{QOHc2}(b) shows the oscillating signal extracted by subtracting a spline background from the raw resistivity. To get a better sensibility, Fourier transforms were made on data extracted using a digital lock-in with a small time constant $\tau=30$~$\mu$s. For clarity, oscillating data extracted using a higher time constant $\tau=100$~$\mu$s are shown in Figure \ref{QOHc2}(b) too. Shubnikov-de Haas oscillations are observed here from 15~T to $\mu_0H_1=35$~T. The corresponding Fourier spectra are shown in Figure \ref{QOHc2}(c) for a large set of small field windows.\cite{note} Figure \ref{QOHc2}(d) presents the field-dependence of the Shubnikov-de Haas frequencies. In all explored field windows between 15 to 30~T, the frequencies $F_\beta$ and $F_\alpha$, which at low field equal $\simeq400$~T and $\simeq1000$~T, respectively, are observed. A progressive frequency change, signature of a continuous Fermi surface modification, occurs within a large field window going from 15 to 30~T, where $F_\beta$ increases while $F_\alpha$ decreases as $H$ increases. The sample-dependent crossover field $\mu_0 H^*\simeq20-25$~T lies in this field window and could possibly be a signature of the associated Fermi surface change. A sudden spectrum modification occurs at $\mu_0 H_{\rho,max}^{LT}=30$~T, at which the frequency $F_\alpha$ reaches $\simeq850$~T, and above which the Fermi surface is reconstructed: for $H_\rho^{LT}<H<H_1=35$~T, $F_\alpha$ remains almost field-independent, we loose the trace of $F_\beta$, and a new frequency $F_\delta\simeq1300$~T appears. For comparison, the frequencies extracted from studies in steady magnetic fields [\onlinecite{altara,jo,aoki12,shishido}] are also plotted in Fig. \ref{QOHc2}(d). An excellent agreement is found between our data and that from Aoki \textit{et al}.\cite{aoki12}, where a similar analysis as here, i.e., with a high number of small field windows, was carried out. Due to the electronic noise of our pulsed field experiment, we were not able to observe here the low frequencies $\eta$, $\gamma$, and $\omega$ found by Aoki \textit{et al}.\cite{aoki12}. Surprisingly, we were able to follow the $\beta$ frequency from 22 to 25~T, even though Aoki \textit{et al.} [\onlinecite{aoki12}] did not observed it. Our data are also in good agreement with that from Altarawneh \textit{et al.},\cite{altara} Jo \textit{et al}.,\cite{jo} and Shishido \textit{et al},\cite{shishido} where Fourier transforms were done on larger field windows. However, a high frequency of $\simeq1500$~T was extracted at $\simeq20$~T by Altarawneh \textit{et al},\cite{altara} but not here nor by Aoki \textit{et al}.\cite{aoki12} As well, we found no trace of the frequency $F_\varepsilon\simeq1300$~T reported by Shishido \textit{et al}.\cite{shishido} above $\mu_0 H=20~T$. In the window 25-30~T, Altarawneh \textit{et al}.\cite{altara} and Jo \textit{et al}\cite{jo} extracted a low frequency of $\simeq250$~T, which was not observed here nor by Aoki \textit{et al}.\cite{aoki12} As well, in the window 30-35~T Altarawneh \textit{et al}.\cite{altara} and Jo \textit{et al}.\cite{jo} found a high frequency of $\simeq1500-2000$~T, which was not observed here nor by Aoki \textit{et al}.\cite{aoki12} The differences between these studies come from the difficulty to extract fine Fourier transform spectra in field windows smaller than a few Shubnikov-de Haas periods. Another difficulty is that the observed frequencies result from the sum or subtraction of harmonic frequencies to the fundamental frequencies, which prevents from extracting real fundamental frequencies. Despite these difficulties, all experimental studies agree on the fact that a magnetic field applied along \textbf{c} induces successive modifications of the Fermi surface in magnetic fields far below $\mu_0 H_1=35$~T, i.e., in the hidden-order phase.

\subsection{Quantum oscillations for $\mathbf{H}\perp\mathbf{c}$}

\begin{figure}[b]
\centering
\includegraphics[width=8cm]{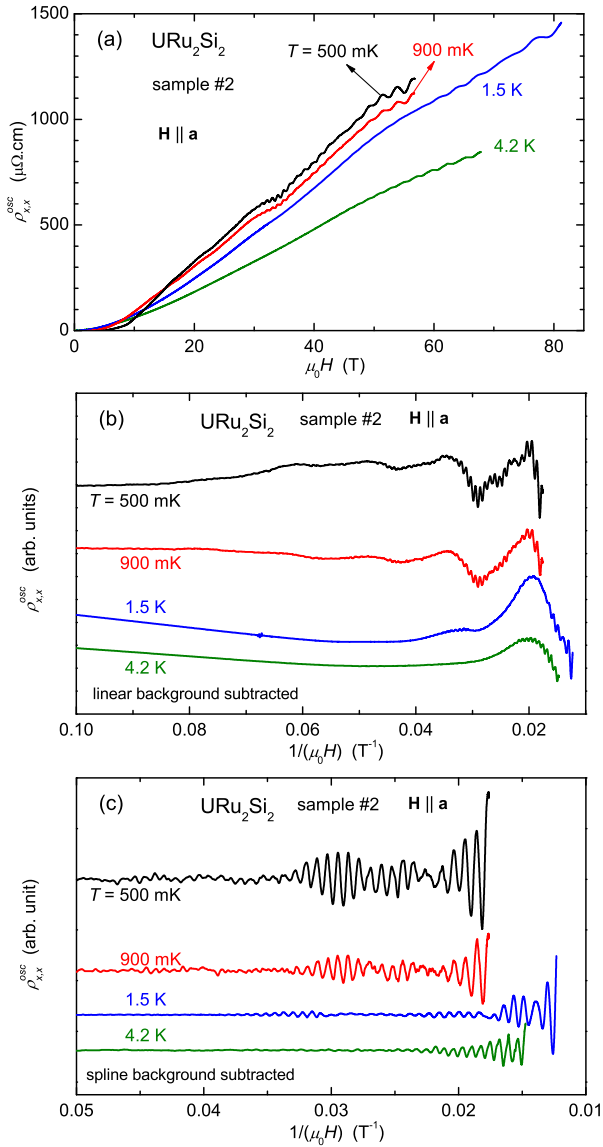}
\caption{For sample $\sharp2$ in $\mathbf{H}\parallel\mathbf{a}$: (a) Transverse resistivity $\rho_{x,x}$ versus $H$, at 500~mK~$\leq T\leq$~4.2~K. (b) Oscillating resistivity $\rho_{x,x}^{osc}$ extracted using a linear background versus $1/H$. (c) Fast oscillating resistivity extracted using a spline background.}
\label{QOHa}
\end{figure}

\begin{widetext}

\begin{figure}[h]
\centering
\includegraphics[width=16.5cm]{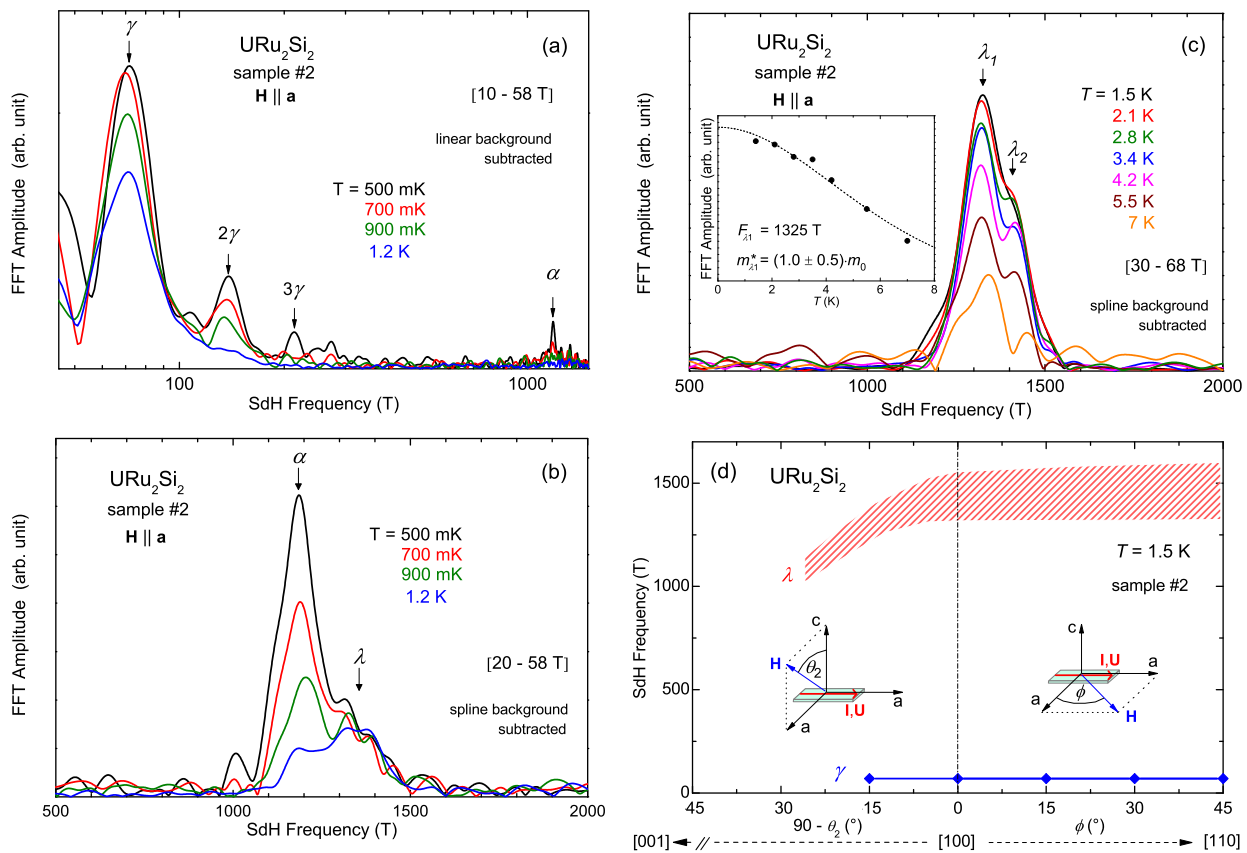}
\caption{For sample $\sharp2$ in $\mathbf{H}\parallel\mathbf{a}$: (a) Fourier spectra of the oscillations extracted using linear backgrounds at 500~mK~$\leq T\leq$~1.2~K. (b) Spectra of the fast SdH oscillations extracted using spline backgrounds at 500~mK~$\leq T\leq$~1.2~K. (c) Spectra of the fast SdH oscillations extracted using spline backgrounds at 1.5~K~$\leq T\leq$~10~K. Inset: plot of the amplitude of $\lambda_1$ versus $T$ and fit of the related mass using the Lifshitz-Kosevich formula. (d) Angular dependence of the SdH frequencies $F_\gamma$ and $F_\lambda$ observed here at $T=$1.5~K. For the $\lambda$ branch, the shaded area indicates the frequency ranges at which the Fourier spectra are enhanced in the panel (c).}
\label{QOHa2}
\end{figure}

\end{widetext}

Figure \ref{QOHa}(a) shows the transverse resistivity $\rho_{x,x}$ of sample $\sharp2$ as function of $\mathbf{H}$ applied along the \textbf{a}-axis, at temperatures from 500~mK to 4.2~K. At $T=500$~mK, the sample is superconducting up to $\mu_0 H_{c,2}\simeq9$~T [defined at a kink in $\rho_{x,x}(H)$], above which $\rho_{x,x}$ increases significantly with $H$, from $\simeq50$~$\mu\Omega$cm at 10~T to 1150~$\mu\Omega$cm at 55~T. No field-induced transition is observed in our resistivity data at $T=1.5$~K and the system remains in the hidden-order phase up to 81~T. This agrees well with the report by Yanagisawa \textit{et al.}. \cite{YANAGISAWA} of a lack of anomaly in the elastic constant in $\mu_0\mathbf{H}\parallel\mathbf{a}$ up to 69~T, at $T=1.5$~K. At $T=500$~mK and above $H_{c,2}$, the non-oscillating part of $\rho_{x,x}$ is almost linear and clearly deviates from the $H^2$-law expected in a one-band Fermi liquid picture. Slow and fast Shubnikov-de Haas oscillations are visible in the raw data up to the highest investigated fields. No change of the SdH frequencies as function of the magnetic field is observed within our experimental resolution, which indicates that the Fermi surface remains unchanged in a high magnetic field up to 81~T applied along $\mathbf{a}$. The slow oscillations shown in Figure \ref{QOHa}(b) were extracted using linear backgrounds. Figure \ref{QOHa2}(a) shows the corresponding Fourier spectra, which exhibit peaks at $F_{\gamma}=70$~T, at its harmonics $F_{2\gamma}=140$~T and $F_{3\gamma}=210$~T, and at $F_\alpha=1185$~T, in good agreement with previous low-field reports. \cite{ohkuni99,hassinger} The fast oscillating signal shown in Figure \ref{QOHa}(c) was extracted using a spline background. The corresponding spectra are shown in Figure \ref{QOHa2}(b): in addition to the main peak at $F_\alpha=1185$~T, a shoulder is attributed to a peak at $F_\lambda\simeq1350$~T. While the intensity of $\alpha$ vanishes rapidly with $T$ due to the high effective mass $m^*_{\alpha}=9.7\cdot m_0$ [\onlinecite{aoki12}], the intensity of $\lambda$ decreases much slower with $T$. Figure \ref{QOHa2}(c) shows, for $1.4\leq T\leq 10$~K, the spectra of sample $\sharp2$ extracted using spline backgrounds. For $T\geq1.4$~K, a higher excitation current allowed to reach a better sensitivity than at subkelvin temperatures (where high excitation currents are prohibited). The spectra show that $\alpha$ has almost totally vanished above 1.5~K and that $\lambda$ survives up to more than 7~K, being split into two frequencies $F_{\lambda_1}\simeq1325$~T and $F_{\lambda_2}\simeq1400$~T. The effective mass $m^*_{\lambda_1}=(1.0\pm0.5)\cdot m_0$ deduced from the temperature dependence of the $\lambda_1$ amplitude is a factor 10 smaller than the effective mass of $\alpha$. \cite{aoki12} Figure \ref{QOHa2}(d) presents the angle-dependence of the Shubnikov-de Haas frequencies of the $\gamma$ and $\lambda$-branches extracted from resistivity measurements in a field applied in the (\textbf{a},\textbf{c}) and (\textbf{a},\textbf{a}) planes (cf. Fig. \ref{figROT1}). A slight increase by $\sim50$~T of $F_{\lambda}$ is observed as the field-direction moves from $\left[100\right]$ ($\phi=0^{\circ}$) to $\left[110\right]$ ($\phi=45^{\circ}$). When the field rotates from $\left[100\right]$ ($\theta_2=90^{\circ}$) to $\left[001\right]$ ($\theta_2=0^{\circ}$), $F_{\lambda}$ decreases more significantly, from $\sim1400$~T at $\theta_2=90^{\circ}$ to $\sim1100$~T at $\theta_2=60^{\circ}$, and its trace is lost at angles $\theta_2<60^{\circ}$. The angle-dependence of the $\lambda$-frequencies, which are observed here in the (\textbf{a},\textbf{a}) plane, and out of the (\textbf{a},\textbf{a}) plane [at angles $(90-\theta_2)$ up to $30^{\circ}$], is compatible with a large and almost spherical Fermi surface similar to the $\alpha$-branch. In agreement with previous reports (see [\onlinecite{aoki12}]), no variation of $F_\gamma$ is observed for the investigated field-directions.

\section{DISCUSSION}

High-quality URu$_2$Si$_2$ samples exhibit a remarkably strong magnetoresistivity inside the hidden-order phase, which is dominated by the orbital effect, as shown by the sample- and angle-dependences of the resistivity. The resistivity of our highest-quality sample increases by three orders of magnitude as a magnetic field applied along the \textbf{a}-axis increases from the low-field range (in the normal state) to 81~T. The high quality of our samples and the high carrier mobility \cite{kasahara} are responsible for this exceptionally large orbital effect. The angle-dependent study of the resistivity shows that the magnetic transitions and the electronic anomalies related to the Fermi surface changes exhibit the same angle-dependence in 1/$\cos\theta$, where $\theta$ is the angle between $\mathbf{H}$ and $\mathbf{c}$, indicating the strong correlation between the Fermi surface and the magnetic polarization induced along the \textbf{c} axis in URu$_2$Si$_2$. A magnetic field applied along the easy magnetic axis \textbf{c} destabilizes the hidden-order phase at $\mu_0 H_1=35$~T, but no anomaly is induced when the magnetic field is applied along the hard axis \textbf{a}, at least up to 81~T. At zero-magnetic-field, the hidden-order phase is thus stabilized by the strong Ising-character of the magnetic properties. At 1.4~K, for $H>H_1$ applied along the \textbf{c}-axis, the resistivity is neither sample-dependent, nor angle-dependent, and has no observable orbital contribution. However, a peculiar sample-dependence of the resistivity in the regime $H_1<H<H_3$ develops at $T=100$~mK, indicating an interplay between the orbital motion of the electrons and the magnetic properties. As observed in the magnetization $M(H)$ (cf. [\onlinecite{devisser86,sugiyama99,scheerer12}]) successive partial polarizations of the 5\textit{f}-electron moments occur at $H_1$, $H_2$, and $H_3$. Observations by Nernst, Hall, and Shubnikov-de Haas effects (cf. [\onlinecite{levallois,altara}]) indicate that these polarizations induce Fermi surface reconstructions, due to reconstructions of the magnetic Brillouin zone, at $H_1$, $H_2$, and $H_3$. We have shown that a magnetic field applied along the \textbf{c}-axis induces anomalies in the orbital contribution to the resistivity at $\mu_0H^*\simeq20-25$~T and $\mu_0H^{LT}_{\rho,max}\simeq30$~T, i.e., at fields well below the destruction of the hidden-order phase at $\mu_0 H_1=35$~T. In the literature, similar anomalies were observed at $\mu_0H^*\simeq23-25$~T not only in the resistivity of other samples\cite{shishido,altara,aoki12} but also in Hall resistivity\cite{shishido,malone} and thermopower\cite{malone,pourret} data. A change of slope of $\rho_{x,x}(H)$ at $\simeq 8$~T is related to a splitting of the $\beta$-branch in Refs. [\onlinecite{hassinger,aoki12}]. In the thermoelectric power, local maxima at $\simeq24$~T and $\simeq30$~T were attributed to the signatures of Lifshitz-transitions.\cite{malone,pourret} The evolution of the Shubnikov-de Haas spectra clearly indicates field-induced Fermi surface reconstructions inside the hidden-order phase, presumably due to successive polarizations of the different Fermi surface pockets. In particular, a Fermi surface reconstruction occurs at $\mu_0 H^{LT}_{\rho,max}=30$~T, at which the resistivity is maximum. We note that the low-temperature magnetization shows no anomaly in the field range 0-35~T (cf. Refs. [\onlinecite{sugiyama99,scheerer12}]). The observed anomalies in the transport properties are thus due to Fermi surface instabilities. The hidden-order parameter and field-induced polarization of the 5\textit{f}-electron magnetic moments are in strong competition, which results in the transition at $\mu_0 H_1=35$~T for $\mathbf{H}\parallel\mathbf{c}$. In Ref. [\onlinecite{scheerer12}], we have shown that the onset of intersite electronic interactions, presumably antiferromagnetic fluctuations, is a precursor of the hidden-order phase. Modeling the interplay between the evolutions of the Fermi surface and the hidden-order and their relation with the magnetic anisotropy is expected to be a key for describing URu$_2$Si$_2$.

A new Fermi surface sheet of frequency $F_\lambda\simeq1400$~T and effective mass $m_\lambda^*\simeq(1\pm0.5)\cdot m_0$ for $\mathbf{H}\parallel\mathbf{a}$ has been observed here by high-field magnetoresistivity experiments. Using the formula, approximated for a spherical Fermi surface, of the Sommerfeld coefficient $\gamma=\sum\limits_{i}\gamma_i\approx\sum\limits_{i}k_B^2Vm_i^*k_{Fi}/(3\hbar^2)$, where $V=49$~cm$^3$/mol is the molar volume, $k_{Fi}=\sqrt{2eF_i/\hbar}$ the wavevector, $F_i$ the SdH frequency, and $\gamma_i$ the contribution to the Sommerfeld coefficient from the Fermi sheet $i$, \cite{hassinger} we estimate the contribution of the new band $\lambda$ by $\gamma_\lambda\simeq0.5$~mJ/molK$^2$, which represents less than 1~$\%$ of the Sommerfeld coefficient $\gamma_{Cp}\simeq 65$~mJ/mol$\cdot$K$^2$ extracted from the specific heat.\cite{maple} Recent cyclotron resonance experiments \cite{tonegawa,tonegawa2013} permitted to report new Fermi surface branches, which had not yet been observed by quantum oscillation techniques. One of these new branches $\kappa$ was found to be particularly heavy, weighting as 30$\%$ of $\gamma_{Cp}$. The new branch $\lambda$ observed here could possibly correspond to one of the four other new and light branches (noted F, G, H, and I) observed by cyclotron resonance in Refs. [\onlinecite{tonegawa,tonegawa2013}]. We note that band calculation models, as those developed in Refs. [\onlinecite{oppeneer11,ikeda12}], might be refined to present the new light band $\lambda$ observed here.

\section{CONCLUSION}

We have performed a systematic investigation of the high-field resistivity of URu$_2$Si$_2$ high quality single crystals in pulsed magnetic fields up to 81~T. As shown by a Kohler plot and by a simple relationship $\mu\propto1/\rho_{x,x}^n(H=0)$ between the carrier mobility for $\mathbf{H}\parallel\mathbf{c}$ and the zero-field resistivity (in the normal state), the non-oscillating low-field magnetoresistivity can be described using a unique relaxation time $\tau$ for all contributing bands. For $\mathbf{H}\parallel\mathbf{c}$, crossovers associated with a kink in $\rho_{x,x}(H)$ at $\mu_0 H^*=20-25$~T and with a maximum of $\rho_{x,x}(H)$ at $\mu_0 H_{\rho,max}^{LT}=30$~T are related to Fermi surface modifications within the hidden-order phase. While $H_{\rho,max}^{LT}$ is almost sample-independent, we find out that $H^*$ is strongly sample-dependent and can be hidden in high-quality crystals where a huge orbital effect contributes to $\rho_{x,x}$. We have established that the low-temperature phase transitions $H_1$, $H_2$, $H_3$ and crossovers $H^*$ and $H_{\rho,max}^{LT}$ are controlled by a 1/$\cos\theta$-law, where $\theta$ is the angle between \textbf{H} and the \textbf{c}-axis. For $\mu_0\mathbf{H}\parallel\mathbf{a}$ up to 81~T, the system remains in its hidden-order state and no Fermi surface change is observed. In this field-configuration, quantum oscillations from a new and almost-spherical branch $\lambda$ of frequency $F_\lambda\simeq1400$~T and effective mass $m_\lambda^*\simeq(1\pm0.5)\cdot m_0$ are observed up to 7~K. As well as the $\alpha$ branch, the $\lambda$ branch is found to be splitted. The work presented here, as an extension to lower temperatures, higher fields, and new field configurations of our work published in Ref. [\onlinecite{scheerer12}] strongly supports that the interplay between the Fermi surface, the magnetic properties, and the hidden-order plays a significant role in URu$_2$Si$_2$. This should be considered for the development of realistic models describing the hidden-order state in URu$_2$Si$_2$.

\section*{Acknowledgements}

We acknowledge K. Behnia, J.-P. Brison, M. Goiran, H. Harima, and J. Levallois for useful discussions, and L. Bendichou, P. Delescluse, T. Domps, J-M. Lagarrigue, J.-P. Nicolin, C. Proust and T. Schiavo for technical support. This work was supported by Euromagnet II via the EU under Contract No. RII3-CT-2004-506239, and by the ERC Starting Grant NewHeavyFermion.


\begin{thebibliography}{9}
\bibitem{note} In the analysis performed here (Figure \ref{QOHc2}), the width of the windows from [13.6~T;18.5~T] to [24.1~T;32.5~T] was adjusted to correspond to seven periods of the main oscillation ($F_\alpha\simeq1000$~T), which is necessary to obtain well-resolved Fourier transforms. At high-field, the Fermi surface is found to be modified in a narrow field window [31.5~T;34.7~T], which corresponds to only four periods of the main observed oscillation ($F_\delta\simeq1300$~T).
\end{thebibliography}
\end{document}